\documentclass{iau_FM}
\usepackage{graphicx}

\title[Models of Stellar Convective Turbulence and Oscillations]{2-D and 3-D Models of Convective Turbulence and Oscillations in Intermediate-Mass Main-Sequence Stars}

\author[Guzik et al.]   
{Joyce A. Guzik$^1$, T.H. Morgan$^{1,2}$, N.J. Nelson$^1$, C. Lovekin$^3$, K. Kosak$^4$, I.N. Kitiashvili$^5$, N.N. Mansour$^5$,
 \and A. Kosovichev$^6$}

\affiliation{$^1$Los Alamos National Laboratory, Los Alamos, NM  USA 87545 \\ email: {\tt joy@lanl.gov}
 \\[\affilskip]$^2$Brigham Young University, Provo, UT  84602 USA 
 \\[\affilskip]$^3$Mount Allison University, Sackville, NB E4L 1H3  Canada
\\[\affilskip]$^4$Florida Institute of Technology, Melbourne, FL  32901 USA
\\[\affilskip]$^5$NASA Ames Research Center, Mountain View, CA 94035 USA
\\[\affilskip]$^6$Physics Department, New Jersey Institute of Technology,
Newark, NJ  07103 USA
}

\pubyear{2015}
\jname{Astronomy in Focus, Volume 2} 
\editors{Piero Benvenuti, ed.}
\begin{document}

\maketitle

\begin{abstract}

We present multidimensional modeling of convection and oscillations in main-sequence stars somewhat more massive than the Sun, using three separate approaches:  1) Using the 3-D planar StellarBox radiation hydrodynamics code to model the envelope convection zone and part of the radiative zone.  Our goals are to examine the interaction of stellar pulsations with turbulent convection in the envelope, excitation of acoustic modes, and the role of convective overshooting; 2) Applying the spherical 3-D MHD ASH (Anelastic Spherical Harmonics) code to simulate the core convection and radiative zone. Our goal is to determine whether core convection can excite low-frequency gravity modes, and thereby explain the presence of low frequencies for some hybrid $\gamma$ Dor/$\delta$ Sct variables for which the envelope convection zone is too shallow for the convective blocking mechanism to drive gravity modes; 3) Applying the ROTORC 2-D stellar evolution and dynamics code to calculate evolution with a variety of initial rotation rates and extents of core convective overshooting. The nonradial adiabatic pulsation frequencies of these nonspherical models are calculated using the 2-D pulsation code NRO. We present new insights into pulsations of 1-2 M$_\odot$ stars gained by multidimensional modeling.

\keywords{stars: convection; stars: oscillations; stars: rotation; {\it Kepler} spacecraft}
\end{abstract}

\firstsection 
              
\section{Introduction and 3-D StellarBox Code}

Motivated by understanding the role of turbulent convection in pulsations of intermediate-mass stars, we used the observed frequency spectra of stars observed by the {\it Kepler} spacecraft to study the transition from solar-like oscillations to $\gamma$-Dor pulsations to $\delta$ Sct pulsations in 1-2 M$_\odot$ stars.

Kitiashvili \etal\ presented results of simulations using the 3-D planar geometry compressible radiative MHD StellarBox code (\cite{wray2015}) at this Focus Meeting.

\section{ASH Models}

We simulated core convection and the adjacent radiative zone of 1.6-2 M$_\odot$ stars using the 3-D spherical hydrodynamics code ASH (Anelastic Spherical Harmonic, see \cite{nelson2013}).
To generate initial conditions, we calculate 1-D models using the MESA evolution code (\cite{MESA}) and created a script to convert the MESA output for input to ASH. The input profiles required include nuclear energy generation, thermal diffusivity, and convection boundaries. The ASH simulations require iterations to gradually lower the viscosity and thermal diffusion to get core convection started.

We added a radiative zone of several pressure scale heights (100 radial zones) above the convective zone including the g-mode propagation region on top of the core. The simulations included 100 radial, 256 angular, and 512 azimuthal points.

\begin{figure}[b]
\begin{center}
 \includegraphics[width=3.4in]{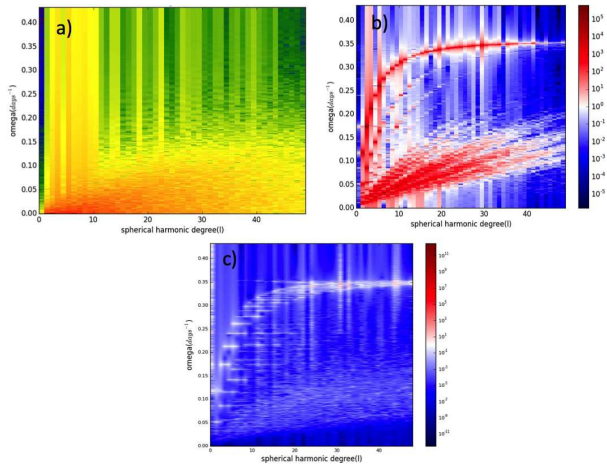} 
 \caption{Synthetic spectrum showing angular frequency $\omega$ vs. spherical harmonic degree $l$ for 1.8 M$_\odot$ ASH zero-age main sequence model without rotation:  a) power in the core convection zone; b) power in radiative zone; c) ratio of power in radiative  to convective zone, indicating strength of coupling to radiative zone of g modes generated in the convective zone.}
   \label{ASH}
\end{center}
\end{figure}

We have first results for gravity-mode excitation for 1.8 M$_\odot$ models on the zero-age main sequence without and with (slow) rotation.  We find that the convective core does excite a wide spectrum of g modes that couple to the radiative zone (Fig. \ref{ASH}), and that may be visible.  Our first models show that the low-degree modes have too low a frequency to be associated with observed $\gamma$ Dor-type g modes, but improvements in the model stratification to better represent the g-mode cavity may change these results.  

 \begin{figure}[b]
\begin{center}
 \includegraphics[width=3.4in]{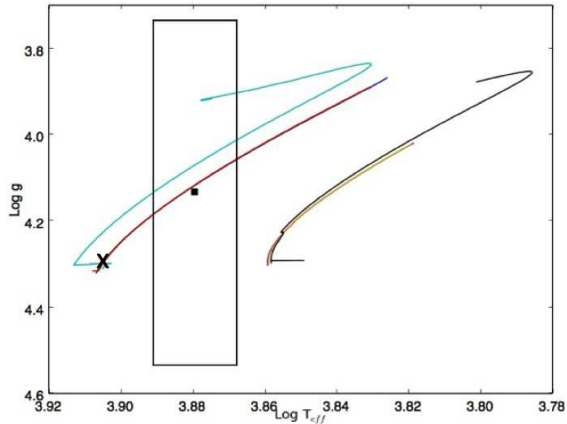} 
 \caption{log surface gravity vs. log T$_{\rm eff}$ for stellar models evolved with ROTORC with and without rotation. The box is  the location of KIC 10451090 using the parameters of the KIC catalog with error bars from \cite{brown2011}.  The $\times$ marks the location of the best-fit model on the 1.7 M$_{\odot}$ evolution track.}
   \label{HRD}
\end{center}
\end{figure}

 \begin{figure}[b]
\begin{center}
 \includegraphics[width=3.4in]{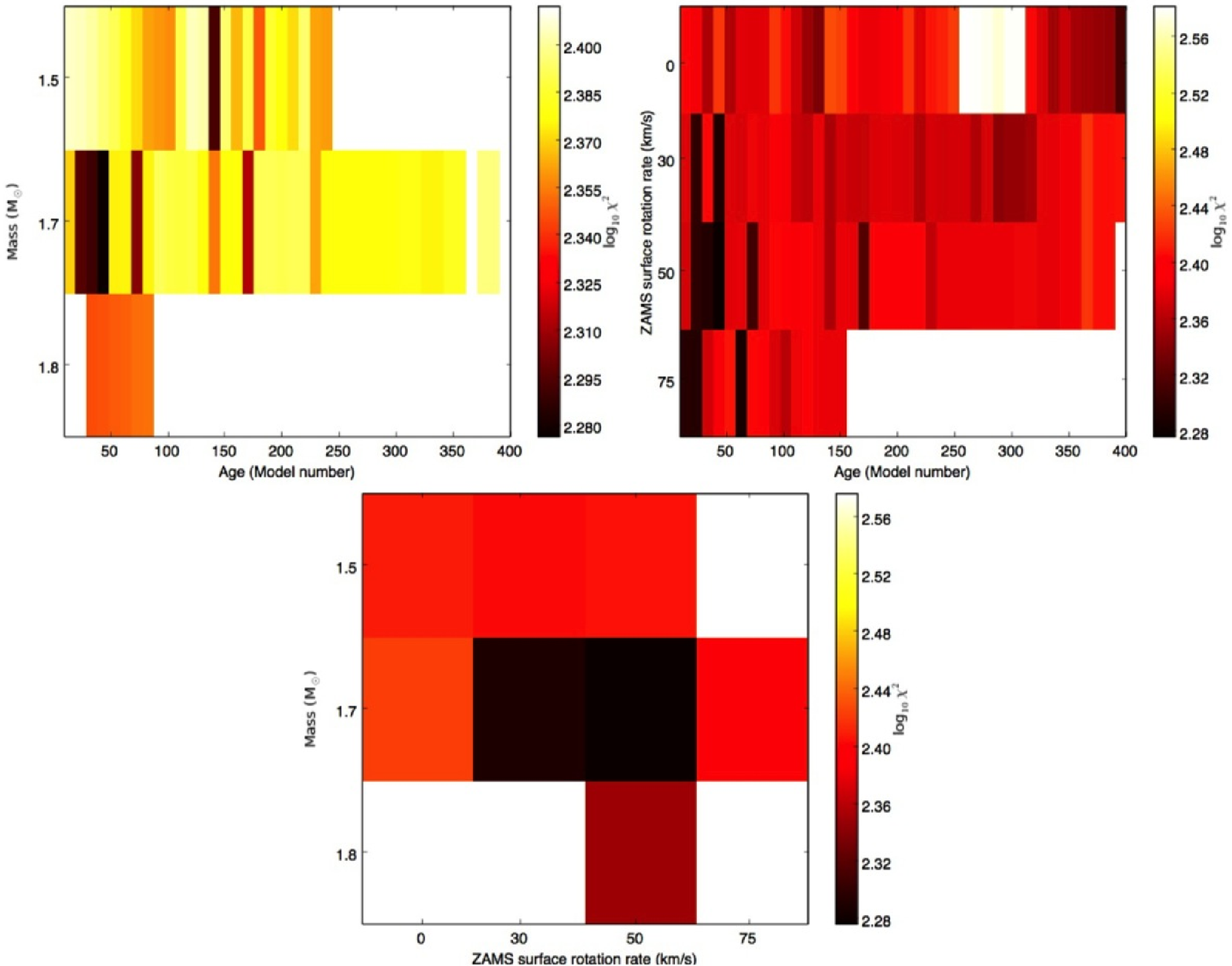} 
 \caption{$\chi^2$ optimization for mass, rotation, and core hydrogen mass fraction (age) for KIC 10451090 using the ROTORC/NRO model grid and match to observed frequencies.}
   \label{opt}
\end{center}
\end{figure}

\section{ROTORC and NRO models}

Stellar models were evolved using ROTORC (\cite{deupree1990, deupree1995}), a fully 2-D stellar structure and evolution code. The models described here have 581 radial zones and 10 angular zones. The equations of stellar structure are solved on the 2-D grid with fractional radius (r/R) and colatitude as the independent variables. 

In rotating models, deformation of the surface is calculated by assuming the surface is an equipotential with the value determined by the value at the equator. The surface in the remaining angular zones is determined to be the radial zone with the total potential closest to the equatorial value. Defining the surface in this way allows the surface parameters to vary as a function of colatitude.

The pulsations were calculated with a 2-D adiabatic pulsation code NRO (\cite{clement1998, lovekin2009}) NRO uses a 2-D finite difference grid to solve the pulsation equations, which allows us to calculate the pulsations for a 2-D stellar model. NRO can include up to 9 angular zones.  The models presented here use three zones; previous work has shown that three zones are sufficient to calculate frequencies even for stars with relatively high rotation rates (\cite{lovekindeupree2008}). The resulting solution is defined at three angular points, which can subsequently be decomposed into individual spherical harmonics. In this work, we have calculated modes with degree $l$ = 0, 1, and 2, with $m$ $\le$ $l$. 

\begin{table}
  \begin{center}
  \caption{ROTORC best-fit $\chi^2$ models.}
  \label{tab1}
  \begin{tabular}{lcccccc}\hline 
KIC ID & Mass M$_{\odot}$ & X$_{\rm core}$ & V$_{\rm eq}$ (km/s) &  L (L$_\odot$) & T$_{\rm eff}$ (K) & R (R$_{\odot}$) \\ 
\hline
5466537 &   1.5 & 0.535  & 21.4  & 5.68 & 7154 & 1.56 \\
8677585 &   1.7 & 0.604  & 0.0 & 9.08 & 7154 & 1.58 \\
10451090 &   1.7 & 0.681  & 47.9  & 8.55 & 7154 & 1.52 \\
 \end{tabular}
 \end{center}
\end{table}

Table \ref{tab1} gives ROTORC/NRO ``best fits'' to frequencies of three {\it Kepler}  stars.  
For the frequency fits, the 50 highest-amplitude significant frequencies calculated from the {\it Kepler} short-cadence data are used.  For KIC 8677585, the frequencies are between 130 and 150 c/d. The other two stars were fit with frequencies between 9 and 40 c/d. 

The fitting algorithm finds the frequency that produces the lowest $\chi^2$ and assigns the observed frequency to that model frequency.  This process is repeated for all other frequencies.  Once all the observed frequencies have been matched to a theoretical frequency, the code calculate a total $\chi^2$ for that model. The model with the lowest $\chi^2$ is considered to be the ``best'' model.  The frequencies are made with a linear combination of three spherical harmonics, $l$ = 0, 1, and 2. The categories are less meaningful as rotation rate increases, but the rotation rates are slow enough that the eigenfunctions should still look like $l$= 0, 1, and 2 modes.

Figure \ref{HRD} shows the evolution in an H-R diagram of 1.5 and 1.7 M$_\odot$  models with and without rotation.  The box shows the location of KIC 10451090 according to {\it Kepler} input catalog parameters, with frequencies of about 40 c/d, and the $\times$ marks the location of the best-fit model on the 1.7 M$_\odot$ evolution track including rotation.  Figure \ref{opt} shows the $\chi^2$ fits for mass, rotation rate, and core hydrogen abundance (representative of age) for this star.  The rotation rates in Fig. \ref{opt} are ZAMS initial rotation rates, and are assumed to be solid body. As the stars evolve, ROTORC conserves angular momentum locally, so the surface rates drop. 

\section{Conclusions}

3-D StellarBox results:  Convective patterns in the envelope of stars more massive than the sun are quite different from solar conditions, and will be important to understand properties of solar-like oscillations and potentially stochastically excited acoustic modes in stars of 1-2 M$_\odot$. 

ASH results: Core convection excites low-degree gravity modes and may help explain observed low frequencies observed in some $\delta$ Sct stars.

ROTORC/NRO results:  2-D rotating stellar evolution and pulsation models show promise for future improved matches observed pulsation frequencies for asteroseismology of intermediate-mass rotating stars.

\acknowledgments

We are grateful for support from the NASA {\it Kepler} Guest Observer Program and Astrophysics Theory programs, Los Alamos National Laboratory, and the DOE Summer Undergraduate Laboratory Internship (SULI) program.

\end{document}